\documentclass[twocolumn,prl,superscriptaddress]{revtex4}

\usepackage{amsfonts,amsmath,amssymb,amsthm}
\usepackage{epic,eepic}
\usepackage[final]{graphicx}
\usepackage{subfigure}

\usepackage{setspace}

\def\bydef{~ \stackrel{\bigtriangleup}{=} ~}

\newcommand{\bei}{\begin{itemize}}
\newcommand{\eei}{\end{itemize}}
\newcommand{\beq}{\begin{equation}}
\newcommand{\eeq}{\end{equation}}
\newcommand{\beqr}{\begin{eqnarray}}
\newcommand{\eeqr}{\end{eqnarray}}
\newcommand{\beqrn}{\begin{eqnarray*}}
\newcommand{\eeqrn}{\end{eqnarray*}}
\newcommand{\brr}{\begin{array}}
\newcommand{\err}{\end{array}}
\newcommand{\bef}{\begin{figure}}
\newcommand{\eef}{\end{figure}}
\newcommand{\al}{\alpha}
\newcommand{\ga}{\gamma}
\newcommand{\om}{\omega}

\newcommand{\sig}{\sigma}

\newcommand{\ty}{\tilde{y}}

\newcommand{\tref}{\tau_r}

\begin{document}

\title{Universal properties of correlation transfer in integrate-and-fire neurons}
\author{Eric Shea-Brown}
\affiliation{Courant Institute of Mathematical Sciences, New York
University, New York, NY 10012} \affiliation{Center for Neural Science,
New York University, New York, NY 10012}
\author{Kre\v{s}imir Josi\'{c}}
\affiliation{Department of Mathematics, University of Houston, Houston,
TX 77204-3008}
\author{Jaime de la Rocha}
\affiliation{Center for Neural Science, New York University, New York,
NY 10012}
\author{Brent Doiron}
\affiliation{Courant Institute of Mathematical Sciences, New York
University, New York, NY 10012} \affiliation{Center for Neural Science,
New York University, New York, NY 10012}

\date{\today}

\begin{abstract}

One of the fundamental characteristics of a nonlinear system is how it
transfers correlations in its inputs to correlations in its outputs.
This is particularly important in the nervous system, where
correlations between spiking neurons are prominent. Using linear
response and asymptotic methods for pairs of unconnected
integrate-and-fire (IF) neurons receiving white noise inputs, we show
that this correlation transfer depends on the output spike firing rate
in a strong, stereotyped manner, and is, surprisingly, almost
independent of the interspike variance. For cells receiving
heterogeneous inputs, we further show that correlation increases with
the geometric mean spiking rate in the same stereotyped manner, greatly
extending the generality of this relationship. We present an immediate
consequence of this relationship for population coding via tuning
curves.

\end{abstract}

\maketitle


Systems with spatially correlated, stochastic forcing provide a rich
field of study, with applications ranging from polymer physics to
population biology~\cite{spatial}.  In neuroscience, much theoretical
work focusses on correlation (or synchrony) between {\it coupled}
neurons receiving uncorrelated external fluctuations~\cite{coup}.
Nevertheless, in sensory systems, external fluctuating signals common
to a group of unconnected cells \cite{common,Doi+04} are a strong
source of correlations. Moreover, signal-independent correlations can
develop within strictly feedforward networks, due to convergent
connections between layers~\cite{ffw,Sha+98}. Furthermore, the
connectivity within biological neural networks is often
sparse~\cite{Hel+00}, suggesting that pairs of neurons rarely influence
one another directly~\cite{Shl+06}. These observations have prompted
experimental and theoretical efforts~\cite{corr,Moreno,BinPow} to
characterize the transfer of correlated input currents to correlated
spikes for pairs of {\it uncoupled} neurons. Despite the
straightforward setting of this problem, few common principles have
emerged.

Recently, we have observed that in response to fluctuating input
currents with \emph{fixed} levels of correlation, the spike (output)
correlation between pairs of neurons increases with the spiking rate
and was approximately independent of the spiking
variance~\cite{delDSJR06}. We provided empirical evidence for this
correlation-rate relationship in {\it in vitro} cortical data and in IF
models with `hard' thresholds, for homogeneous cell pairs firing at low
to medium rates. In this letter, we develop an analytical approach that
both explains and greatly extends these observations. Specifically, we
obtain closed-form expressions relating output correlations to firing
(spiking) rate, show that a refractory period implies the existence of
optimal firing rates for correlation transfer, and show that
correlation transfer depends approximately {only} on the geometric mean
of the output firing rates for heterogeneous cell pairs.  We
demonstrate extensions to the quadratic integrate and fire model and to
a broader analysis of the {\em in vitro} data.  These results enable us
to develop the first consequences of the correlation-rate relationship
for cortical coding, where populations of heterogeneously tuned neurons
relay information about the sensory environment.

\begin{figure}[htb]
 \vspace{-.25cm}
     \begin{center}
 {\includegraphics[width=3.3in]{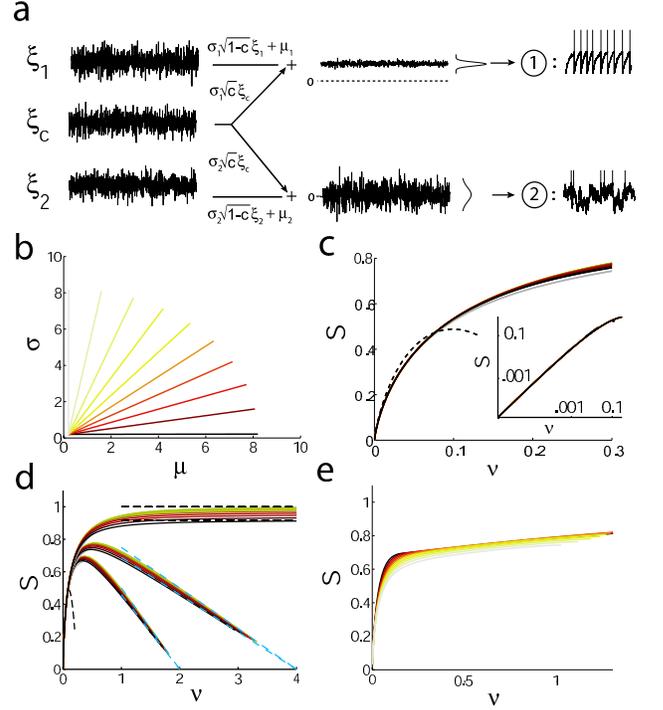}}  \vspace{-.5cm}
     \end{center}
 \caption{a)  Schematic of two cells receiving heterogeneous, correlated white noise inputs.
 b)  Rays with different $\beta=\mu/\sigma$; greyscale (color
 online) applies to the following panels.  c) $S$ vs. $\nu$ increasing
 along the different rays for a limited range of rates for LIF cells, $\tref=0$.
 Here and for all figures with LIF model: $V_{T}=1$, $V_{R}=0$.  Kramer's approximation shown as dotted line.  Inset, on log-log axes. d) As (c), for wider range of rates.  Dotted
lines show Kramer's approximation at low rates and large $\mu$, $\sig$
approximations at high rates.  Sets of curves, from top to bottom:
$\tref=0$, $\tref=0.25$, $\tref=0.5$.  d) Similar, but for QIF model
with $\tref=0$, $V_{T}=10$, $V_{R}=-10$.}
 \label{f.homogeneous}
 \end{figure}

\emph{Model} -- The model consists of two IF neurons receiving common
$\xi(t)$ and independent $\xi_{i}(t)$ white noise inputs
(Fig.~\ref{f.homogeneous}a):\vspace{-.6cm}
\begin{equation} \label{e.lan} \tau V'_i
= f(V_i) +\overbrace{\mu_i + \sigma_i \sqrt{\tau} \left[
\sqrt{1-c}\,\xi_{i}(t)+\sqrt{c}\xi(t) \right]}^{I_i(t)} \,,
\end{equation}
$i = 1,2$.  Here, $V_i$ denotes the membrane voltage of cell $i$,
$\tau$ is the membrane time constant, and $f(\cdot)$ defines the
subthreshold dynamics.  We mostly take $f(V_i)=-V_{i}$ (leaky IF), but
in one section use $f(V_i)=V_{i}^2$ (quadratic IF).  We adopt the
standard threshold-spike-reset condition: $V_{i}(t^{+})=V_{R}$ whenever
$V_{i}(t) = V_T$; at such times, a spike is emitted, after which the
voltage is held at $V_R$ during the \textit{refractory period} $\tref$.
Throughout, we set $\tau = 1$ (we report rates in units of
$\tau^{-1}$).  Of interest are the normalized output spike trains
$y_{i}(t)=\sum_{i}\delta(t-t_{i}^k)$, where $t_{i}^k$ is the $k^{th}$
spike time of the $i^{th}$ neuron.  The firing rate of the $i^{th}$
cell is $\nu_i= \langle y_i(t) \rangle$.

The input current $I_i(t)$ to cell $i$ has a constant (DC) component
$\mu_i$ as well a stochastic component with amplitude $\sig_i$
(Fig.~\ref{f.homogeneous}a). The stochastic component to each cell is
the sum of two independent gaussian processes, $\xi_i(t)$ and $\xi(t)$,
with $\langle \xi_i(t) \xi_j(t') \rangle = \delta_{ij}\delta(t-t')$ and
$\langle \xi_i(t) \xi(t') \rangle = 0$ ($i,j=1,2$); note that $\xi(t)$
is common to both cells.  The weights $\sqrt{1-c}$ and $\sqrt{c}$
result in a correlation coefficient $c$ between $I_1(t)$ and $I_2(t)$.
The spike correlation coefficient, $\rho$, is computed
\cite{Bair,delDSJR06,Moreno,Kohn} by integrating the spike auto- and
cross-correlation functions $C_{ij}(\tau)= \langle y_i(t)y_j(t+\tau)
\rangle - \nu_i \nu_j$:
\begin{equation}
\rho =
\frac{\int_{-\infty}^{\infty}C_{12}(\tau)d\tau}{\sqrt{\int_{-\infty}^{\infty}C_{11}(\tau)d\tau}\sqrt{\int_{-\infty}^{\infty}C_{22}(\tau)
d\tau}}.\label{e.rho}
\end{equation}

\emph{Linear response calculations} -- Following methods
of~\cite{Doi+04}, we derive an analytic expression for $\rho$ valid to
linear order in $c$; simulations of Eq.~\eqref{e.lan} verify that these
methods produce good approximations for $c$ up to $\approx
0.3$~\cite{delDSJR06}, and were repeated to the same effect for the
cases studied here (data not shown).

The cross spectrum of the spike trains $y_1(t)$ and $y_{2}(t)$ is
$P_{12}(\omega)= \langle \ty^*_1 (\om) \ty_2 (\om) \rangle$, where the
Fourier transform $\ty_i=\frac{1}{\sqrt{T}} \int_0^T e^{i \om t}
(y_i(t)-\nu_i) dt$. Treating the common term $\sqrt{c}\xi(t)$ as a
perturbation, we use the linear approximation~\cite{Doi+04}:
\begin{equation} \tilde{y}_{i}(\omega) =
\ty_{i,\mu_i,\sig_i \sqrt{1-c}}(\omega) + \sigma_i \sqrt{c}
A_{\mu_i,\sig_i \sqrt{1-c}}(\omega) \tilde{\xi}(\omega) + {\cal{O}}(c)
. \label{e.linresp1}
\end{equation}
Here $\ty_{i,\mu_i,\sig_i \sqrt{1-c}}(\omega)$ is the Fourier transform
of a spike train obtained from a neuron driven only by the ``private''
input $\mu_i + \sigma_i \sqrt{1-c} \xi_i(t)$ (i.e., in the
\emph{absence} of a common stochastic input), and $A_{\mu_i,\sig_i
\sqrt{1-c}}(\omega)$ is the susceptibility function of such a
neuron~\cite{Ris89}.

Inserting Eq.~\eqref{e.linresp1} into the definition of the cross
spectrum we obtain:
\begin{eqnarray}
P_{12}(\om) &=& c \sigma_1 \sigma_2  A^*_{\mu_1,\sig_1}(\omega)
A_{\mu_2,\sig_2}(\om) P_\xi(\om) + {\cal{O}}(c^{3/2}) \, \label{e.S12}
\end{eqnarray}
where $P_\xi (\omega) = \langle \tilde{\xi}^{*} (\omega)
\tilde{\xi}(\omega)  \rangle$  $=1$.  Here we used $\langle
\ty^*_{i,\mu_i,\sig_i \sqrt{1-c}} \, \ty_{j,\mu_j,\sig_j \sqrt{1-c}}
\rangle=\langle \ty^*_{i,\mu_i,\sig_i \sqrt{1-c}} \,
 \tilde{\xi} \rangle =0$, $i,j=1,2$.  We also used the fact that the ${\cal{O}}(c)$ terms in Eq.~\eqref{e.linresp1} arise from
$\xi$, so are statistically independent of the $\ty_{i,\mu_i,\sig_i
\sqrt{1-c}}$.  Moreover, we evaluate the susceptibility terms at $c=0$,
which is valid to the order of error shown.

The Wiener-Khinchine Theorem 
yields $P_{12}(0)$ = $\int_{-\infty}^{\infty}C_{12}(t) dt$; also, the
susceptibility to DC input is $A_{\mu_i,\sig_i}(0)=\frac{d
\nu_i}{d\mu_i}$. Thus, substituting Eq.~\eqref{e.S12} into Eq.~\eqref{e.rho} yields to lowest order in $c$:
    \beq
    \rho  = c \sigma_1 \sigma_2  \frac{ \frac{d \nu_1}{d
\mu_1} \frac{d \nu_2}{d \mu_2}  }
    { CV_1 CV_2  \sqrt{\nu_1 \nu_2} } \;,  \label{e.heterowhite}
    \eeq
where we have additionally used the result for renewal point processes
$\int_{-\infty}^{\infty}C_{ii}(\tau)d\tau = CV_i^2  \nu_i$; here,
$CV_i$ is the coefficient of variation of the inter-spike interval
distribution \cite{Cox+66}.

\emph{Dependence of correlations on firing rate.} -- 
We first assume that the statistics of inputs to cells $i$ and $j$ are
identical ($\mu_i=\mu_j \equiv \mu $ and $\sig_i=\sig_j \equiv \sig $),
giving
    \beq
    \rho  = c \sigma^2 \frac{ \left( \frac{d \nu}{d
\mu} \right)^2 }
    { CV^2  \nu }  \stackrel{\bigtriangleup}{=} c S(\mu,\sigma)
    \label{e.rhosimple} \;.
    \eeq
Here $S$ is the \textbf{\textit{correlation
susceptibility}}~\cite{delDSJR06} between the input current correlation
$c$ and the spike (output) correlation $\rho$.

We explain three key features of Eq.~\eqref{e.rhosimple}
: {\it i}) $S$ increases with firing rate $\nu$ for low to moderate
rates. {\it ii}) $S$ is {approximately} a function of $\nu$
\textit{alone}, independent of the precise values of $\mu$ and $\sigma$
or the CV. At low rates this relationship becomes exact. {\it iii}) For
large rates: $S$ tends to a constant near 1 if $\tref=0$ and tends to
zero if $\tref>0$.

To study the low rate case we treat the voltage of each cell as a
particle in the potential $\phi(V)=(V-\mu)^2/2$. Low rates are obtained
only when $\mu<V_T$ and $0<\sigma \ll 1$.  In this limit, Kramer's
theory can be applied to calculate the terms in $S$, by computing the
mean ($m$) and the variance ($\mbox{var}$) of hitting times of the
threshold voltage $V_T$ for particles starting at the bottom of the
wells, and by noting that $\nu=(m+\tref)^{-1}$ and $CV^2=\nu^2
\mbox{var}$. Asymptotic results \cite{Bender} give, to lowest order in
$\sigma$,
    \beq
\nu= \frac{\al}{\sqrt{\pi}} \exp \left( - \al^2 \right) \bydef g(\al)
\;, \label{e.nukramers}
    \eeq
and $ \frac{ d \nu}{d \mu} = \frac{\nu}{\sig} \left( 2\al -
\frac{1}{\al} \right),$ where $\al=\frac{V_T - \mu}{\sig}$.
Moreover, $CV^2=1$ to lowest order in $\sigma$, as the threshold
crossing times are exponentially distributed in this limit.  Therefore,
here we have $ S \approx \nu \left( 2 \al - 1/\al \right)^2. $
 Eq.~\eqref{e.nukramers} may be uniquely inverted for a range of {sufficiently}
small $\sigma$ (large $\al$) to yield $\al = g^{-1}(\nu)$.  For small
$\nu$ we obtain:
    \beq
S(\mu,\sig) \approx \hat{S}(\nu)= \nu \left( 2 g^{-1}(\nu) -
\frac{1}{g^{-1}(\nu)} \right)^2,
    \eeq
so that asymptotically $S$ is a function of $\nu$ \textit{alone} (
Fig.~\ref{f.homogeneous}c).

Asymptotics for large firing rate $\nu$ are obtained from exact
expressions for the firing rate and variance~\cite{LindnerThesis}.
Using asymptotic approximations,
we obtain for fixed $\sigma$ and large $\mu$
$$
S(\mu,\sigma) \approx \frac{(V_T-V_R)}{\mu \tref + (V_T-V_R)} \;.
$$
Thus,
the correlation susceptibility $S$ 
approaches 0 in the large $\mu$ limit, as long as $\tref > 0$.  In the
case $\tref = 0$, we obtain $S \rightarrow 1$ as $\mu \rightarrow
\infty$ (Fig.~\ref{f.homogeneous}d).

More generally, if $\mu$ and $\sigma$ diverge jointly so that $\mu/
\sig = \beta$ is constant (i.e., along a ray in
Fig.~\ref{f.homogeneous}b), we obtain $ S(\mu=\beta\sigma,\sigma)
\approx \frac{  K_1(\beta) (V_T-V_R)}{
 K_2(\beta)(V_T-V_R) + \sigma \tref }.$  Here, $K_1(\beta) = \frac{
\left( -2/\sqrt{\pi}
 + 2 \beta e^{\beta^2} \mbox{erfc}(\beta) \right)^2 }{2e^{\beta^2}\int_{\beta}^{\infty} \;
 e^{x^2} \mbox{erfc}^2{(x)} dx}$ and $K_2(\beta) = \sqrt{\pi}e^{\beta^2}\mbox{erfc} (\beta
 )$.   Using asymptotic expressions for $\nu$ at large
$\mu$ and $\sigma$, we obtain
    \beq
S(\mu,\sig) \approx \hat{S}(\nu)= \frac{K_1(\beta)}{K_2(\beta)} \left(1
- \tref \nu \right) \;.
    \eeq
The constant ${K_1(\beta)}/{K_2(\beta)}$ increases from $0.918$ to 1,
as $\beta$ increases from 0 to $\infty$.  Therefore, if $\tref=0$,
$S(\nu)$ is asymptotic to values in the interval $[0.918,1]$ for large $\nu$, as
shown in Fig.~\ref{f.homogeneous}d. Furthermore, if $\tref \neq0$ then
$S(\nu)$ tends to zero at $\nu = \tref^{-1}$, with slope in $[-1,
-0.918]$.


In summary, we have shown that, at low rates, $S$ is asymptotically a
function of $\nu$ \emph{only}, regardless of specific values of $\sig$
and $\mu$ and hence of the $CV$.  At high rates, if $\tref=0$ then $S$
is bounded in a neighborhood of 1; if $\tref>0$ then $S$ tends to zero
within a narrow ``cone" at rates near the limit $\nu = \tref^{-1}$.  It
follows that a refractory period implies that $S$ obtains a maximum at
a particular firing rate, at which the neuron is ``best-tuned" to
transfer input correlations to output correlations.  For intermediate
rates, numerical evaluations of Eq.~\eqref{e.rhosimple} show that $S$
remains an approximate function of $\nu$ alone, a fact that greatly
simplifies the statistical description of the LIF model. We note that
these findings hold more generally: Fig.\ref{f.homogeneous}e shows
analogous results for the quadratic integrate and fire model,
demonstrating that our results depend on neither the linear
subthreshold dynamics nor the ``hard'' threshold condition of the LIF
model.



\begin{figure}[htb]
 {\includegraphics[width=3.3in]{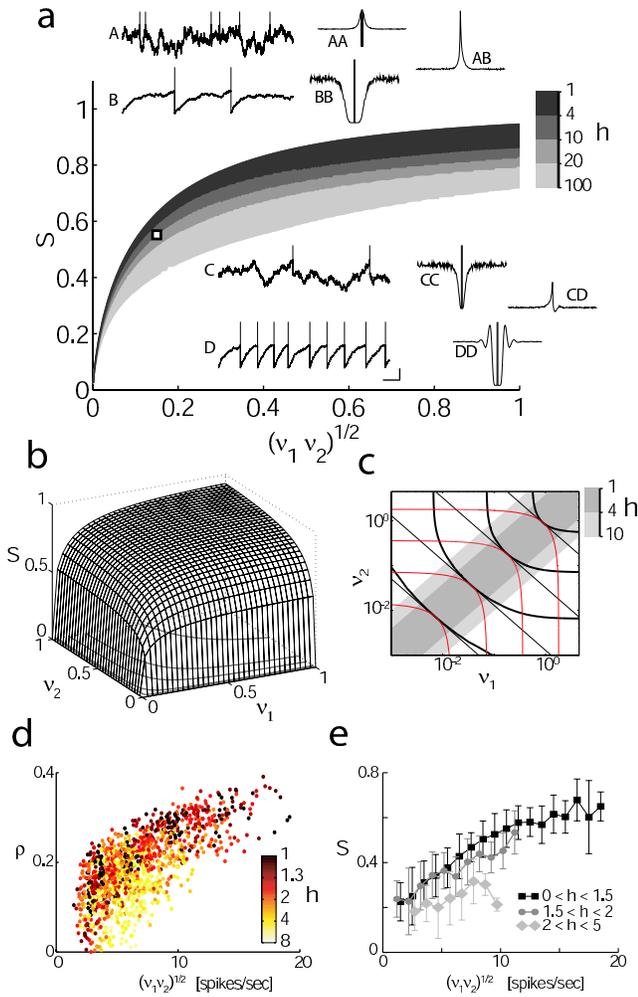}}

 \caption{(a) $S$ vs. $\sqrt{\nu_1 \,\nu_2}$, from
 Eq.~\eqref{e.heterowhite}.  Input parameters
 $\{\mu_1,\sigma_1,$ $\mu_2,\sigma_2\}$ were sampled over a 4-D cube
($\mu_{i,j} \in [.2,8.2]$, $\sig_{i,j} \in [.2,8.2]$, grid spacing
0.01-.05, with denser grid for $\mu,\sig<1.6,2.2$). Overlapping regions
shaded from dark to light contain all points with levels of
$h=\nu_1/\nu_2$ varying over increasingly broad ranges. Square marks
value for the pairings AB,AC,BD,CD shown in the inset (voltage traces
and auto- and cross-correlation functions plotted); see text.
Horizontal scale bar: $2\tau$ for voltage traces, $4\tau$ for
correlation functions. Vertical scale bar: $.25$ for correlation
functions, $.5$ for traces B,D, $1$ for traces A,C. (b) Surface/contour
plot of $S$, where $\sig=\mu$ for both cells. (c) Contour plot on
log-log axes; see text. (d,e)  $S$ vs. $\sqrt{\nu_1 \,\nu_2}$ for {\it
in vitro} cortical experiments; (d), raw data 1269 pairwise comparisons
obtained from 20 cells using $c=0.5$, (e), population average and
standard deviation to show trends.} \label{f.heterogen}
 \end{figure}

\emph{Effects of heterogeneity.} -- We return to the general
Eq.~\eqref{e.heterowhite}.  In neuroscience, several studies plot the
output correlation of two cells versus the geometric mean of their
firing rates $\sqrt{\nu_1 \nu_2}$~\cite{BinPow,Kohn,Bair}.
Fig.~\ref{f.heterogen}a shows the corresponding predictions of
Eq.~\eqref{e.heterowhite}, as the input parameters
$\{\mu_1,\sigma_1,\mu_2,\sigma_2\}$ are sampled over a wide range,
producing very different combinations of $\nu_i, CV_i$, $i=1,2$ (here
and below we take $\tref=0$). Remarkably, when $\rho$ is plotted vs.
$\sqrt{\nu_1 \nu_2}$, we see approximately the \textit{same}
correlation-rate relationship as for the homogeneous case.  At low
rates, this relationship becomes almost independent of the ratio
$h=\nu_1/\nu_2$ ($\nu_1>\nu_2$) between the cells' rates; at higher
rates, it depends only weakly on $h$. Thus, in the limit of small $c$
we have:
    \beq
    \rho  =  c \sqrt{ S(\mu_i,\sigma_i) S(\mu_j,\sigma_j) } \approx c \sqrt{ \hat{S}(\nu_i) \hat{S}(\nu_j) } \approx c \hat{S}(\sqrt{\nu_i \nu_j}).  \label{e.rhoShet}
    \eeq
Here, the first approximation follows from our results for the
homogenous case, and the second follows empirically from
Fig.~\ref{f.heterogen}a.

To see how surprising this is, consider the four voltage traces in
Fig.~\ref{f.heterogen}a, each of which displays either a high or low
CV, and a high or low rate ($\nu=$0.47 or 0.047, resp.). Nevertheless,
any combination of these cases that gives the {\it same} geometric mean
(AB, AC, BD, CD, $\sqrt{\nu_1 \nu_2}=0.15$) gives the \textit{same}
value of $S=0.55$, within less than $1\%$ (square on graph). This is
despite the strongly distinct structure of the correlation functions
for these different pairings (AB, CD shown). Other striking predictions
similarly follow from Eq.~\eqref{e.rhoShet}: for example, simply adding
a DC current to one of a pair of cells (thus increasing, e.g., $\nu_i$)
should increase their spike correlation.

Why does Eq.~\eqref{e.rhoShet} hold? Fig.~\ref{f.homogeneous} shows
that $S(\mu_i,\sigma_i) \approx \hat{S}(\nu_i)$, with exact agreement
at low rates.  The approximation $\sqrt{ \hat{S}(\nu_i) \hat{S}(\nu_j)
}$ $ \approx \hat{S}(\sqrt{\nu_i \nu_j})$ would be exact if
$\hat{S}(\nu_i)$ was a power law, $\hat{S}(\nu_i) \propto \nu_i
^{\ga}$, and Fig.~\ref{f.homogeneous}c (inset) shows that this is an
excellent approximation at low rates.  Hence Eq.~\eqref{e.rhoShet}
holds precisely at low rates, so that the different areas in
Fig.~\ref{f.heterogen}a converge. To explore when the approximation
$\sqrt{ \hat{S}(\nu_i) \hat{S}(\nu_j) }$ $ \approx \hat{S}(\sqrt{\nu_i
\nu_j})$ breaks down, we study the special case where the ratio
$\beta_i=\sig_i/\mu_i=1$ is {identical} for both cells
(Fig.~\ref{f.heterogen}b). Fig.~\ref{f.heterogen}c shows level sets of
this surface on log-log axes (thick lines), which almost superpose with
level sets of $\sqrt{\nu_i \nu_j}$ (thin black lines) for low rates, or
for values of $h$ reasonably close to 1 (shaded areas). For larger
rates, or greater heterogeneity, these level sets start to diverge.
However, even then the disagreement is quite mild: for $h$ up to 10,
values of $S$ along level sets of $\sqrt{\nu_i \nu_j}$ do not vary by
more than $\approx 10 \%$, because $S$ varies only weakly with large
rates. This agreement stands in contrast to that between level sets of
$S$ and of the arithmetic mean $(\nu_1+\nu_2)/2$, shown as grey lines
(red online), indicating that the choice of the geometric mean is
somewhat special in capturing the rate-dependence of $S$.
Fig.~\ref{f.heterogen}d shows that this relationship holds not only for
the LIF model, but also for our {\it in vitro} experiments. Here, pairs
of cortical cells in mouse brain slices were injected with correlated,
noisy, low-pass filtered currents, replicating the configuration of
Fig.~1a (see~\cite{delDSJR06} for experimental methods).  Output spikes
were recorded via whole-cell patch clamp, and $\rho$ was computed from
paired spike trains as described above.




 \begin{figure}[htb]
 {\includegraphics[width=3.3in]{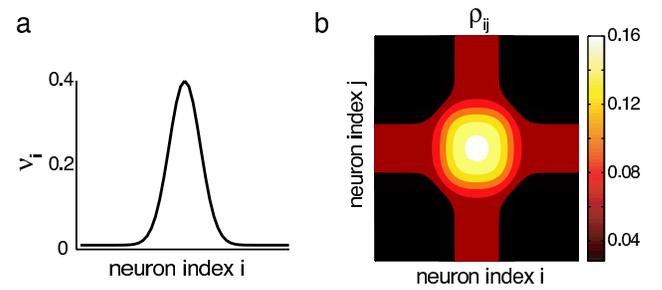}}
\caption{(a) Firing rate in response to a given stimulus for a
population of neurons with tuning curves evenly distributed across the
stimulus ($\theta$) space. (b) The resulting crosscorrelations
$\rho_{ij}$ for this stimulus, from Eq.~\eqref{e.heterowhite}.}
\label{f.tuning}
\end{figure}

For an illustration of the functional consequences of the
rate-correlation relationship for heterogeneous cells, consider a
population of neurons whose rates $\nu_i$ are tuned to a sensory
variable $\theta$ as in Fig.~\ref{f.tuning}a~\cite{Hub+62,corrcode}.
The correlation $\rho_{ij}$ between pairs of these cells approximately
depends on their geometric mean rate $\sqrt{\nu_1 \nu_2}$ according to
Eq.~\eqref{e.rhoShet}, producing the correlation tuning shown in
Fig.~\ref{f.tuning}b (input correlations are taken to be constant:
$c_{ij} \equiv c$).  Thus, the fact that only neurons with similar
tuning display substantial correlation, as in
empirical~\cite{Kohn,Zoh+94} and theoretical~\cite{corrcode} studies,
follows directly from our results, without requiring a separate
assumption that the input correlation is itself tuned.  We also predict
stronger correlations for neurons with ``preferred'' tuning to a
stimulus, as in~\cite{Kohn}. Such an extension of stimulus selectivity
from rate~\cite{Hub+62} to correlation tuning has broad and novel
implications for information {\it encoding} in spiking neurons.
Further, correlations in output spike trains are effective in
propagating activity downstream~\cite{ffw}, thereby facilitating the
{\it decoding} of correlation-based information.



Our methods can be extended: Eq.~\eqref{e.heterowhite} holds for any
cell or cell model producing spikes as a renewal process, and the
preceding forms of this expression are valid even for nonrenewal
spiking (enabling, e.g., analysis of temporally correlated
inputs~\cite{Moreno}). Moreover, our formulas and their reduction to a
compact correlation-rate relationship may be a critical component for
mean field models of sparsely coupled neural {populations}, where
treatment of second-order statistics has previously been restricted to
spike train variance (e.g.,~\cite{mfield}). A mean field framework that
rigorously includes correlations would connect neural population
dynamics~\cite{coup,ffw,mfield} with the emerging role for correlations
in sensory coding.



We acknowledge Steve Coombes and Dan Tranchina for their critical reading and helpful comments, and the support of a BWF Career Award and NSF MSPRF (ESB), grants DMS-0604229 and Texas ARP/ATP (KJ), HSPF (BD), and the Spanish MEC (JR).

\end{document}